\documentclass[11pt,notitlepage,amsmath,amssymb,aps,prb]{revtex4-2}

\usepackage[utf8]{inputenc}
\usepackage{graphicx}
\usepackage{dcolumn}
\usepackage{bm}
\usepackage{xcolor}
\usepackage{svg}
\usepackage{subcaption}
\usepackage[compatibility=false]{caption}
\usepackage{floatrow}
\usepackage{soul}
\usepackage{verbatim}
\usepackage{subcaption}
\usepackage{csquotes}
\usepackage[normalem]{ulem}
\usepackage{lipsum} 
\usepackage[colorlinks=true, linkcolor=blue, citecolor=blue, urlcolor=blue]{hyperref}
\hypersetup{
    colorlinks=true,
    linkcolor=blue,
    filecolor=magenta,
    urlcolor=cyan,
    pdftitle={Paper Overview},
}

\begin{document}


\begin{titlepage}

\title{Effect of static magnetic island on ITG of ADITYA-U tokamak}

\author{Vibhor Kumar Singh$^{1,*}$}
\author{Amal R Biju$^{1}$}
\author{Jaya Kumar Alageshan$^{1}$}
\author{Kaushalender Singh$^{2,3}$}
\author{Deepti Sharma$^{2}$}
\author{Joydeep Ghosh$^{2,3}$}
\author{Nishant Sirse$^4$}
\author{Abhijit Sen$^{2,3}$}
\author{Sarveshwar Sharma$^{2,3}$}
\author{Manjunatha Valmiki$^5$}
\author{Sandeep Agrawal$^5$}
\author{Sanjay Wandhekar$^5$}
\author{Animesh Kuley$^{1,*}$}

\affiliation{$^{1}$Department of Physics, Indian Institute of Science, Bangalore, INDIA - 560012}
\affiliation{$^{2}$Institute for Plasma Research, Bhat, Gandhinagar 382428, India}
\affiliation{$^{3}$Homi Bhabha National Institute, Anushaktinagar, Mumbai, 400094, India}
\affiliation{$^{4}$Institute of Science and Research and Centre for Scientific and Applied Research, IPS Academy, Indore-452012, India}
\affiliation{$^{5}$Centre for Development of Advanced Computing, Pune-411005, India}

\email{vibhorsingh@iisc.ac.in, akuley@iisc.ac.in}

\begin{abstract}
\noindent Magnetic islands play a crucial role in regulating plasma confinement in tokamaks by interacting with micro-instabilities, such as the ion temperature gradient (ITG) mode. This work presents a detailed investigation of the effects of static magnetic islands on ITG instability, relevant to the ADITYA-U tokamak, using the Global Gyrokinetic Code in Cylindrical Coordinates (G2C3), a particle-in-cell (PIC) framework that employs a neural-network-assisted projection scheme. A two-phase simulation strategy is adopted. In the first phase, static magnetic islands with mode numbers (m, n) = (2, 1) and (3, 1) are introduced by perturbing the equilibrium magnetic flux functions. Particle dynamics within these modified topologies result in the flattening of plasma density profiles in the island regions, confirming island formation and its impact on the equilibrium profiles. In the second phase, the flattened profiles serve as new equilibria for linear electrostatic gyrokinetic simulations with adiabatic electrons, enabling the study of the modified ITG behavior. Magnetic islands significantly restructure the ITG mode, producing a spatial redistribution of potential fluctuations within and around the island region. Moreover, as the island width increases, the growth rates of different toroidal ITG modes converge, suggesting a universal stabilization trend. A comparison between the (2,1) and (3,1) islands indicates that higher-q islands lead to a more spatially extended ITG mode structure, reflecting the longer magnetic connection lengths and weaker curvature drive at outer flux surfaces. These results demonstrate the pivotal role of island-induced equilibrium modifications in determining ITG stability and mode structure in tokamak plasmas.
\end{abstract}



\maketitle

\end{titlepage}

\section{Introduction}\label{sec:introduction}

\noindent Magnetic islands (MIs) are coherent structures \cite{F.L.} in magnetized plasmas, characterized by localized perturbations of the magnetic field arising from magnetic reconnection. They emerge in both laboratory and astrophysical plasmas and play an important role in plasma dynamics and energy redistribution processes \cite{Yamada}. In magnetically confined fusion devices such as tokamaks, magnetic islands are of particular interest because of their strong influence on plasma confinement and stability. Their formation and evolution are typically governed by tearing mode instabilities \cite{Harold,hender2007mhd}, which disrupt nested magnetic flux surfaces and modify the magnetic topology, thereby altering plasma profiles and confinement properties \cite{Fitzpatrick}. The presence of magnetic islands can lead to confinement degradation through the flattening of temperature and density profiles and by facilitating secondary instabilities \cite{choi}. Consequently, a comprehensive understanding of magnetic islands and their interaction with plasma fluctuations is crucial for improving performance in fusion devices.

\noindent Magnetic islands formed due to tearing instabilities also play a crucial role in shaping the plasma equilibrium by altering local gradients and field-line connectivity\cite{Wesson2011}. Understanding how such equilibrium modifications influence microinstabilities, particularly ion-temperature-gradient (ITG) modes, is therefore essential for interpreting confinement behavior in tokamaks. Inspired by the work of Jiang \cite{Jiang2014}, who incorporated static magnetic islands into the Gyrokinetic Toroidal Code (GTC) to investigate their impact on ITG instabilities, we extend this line of investigation to the ADITYA-U configuration. Their study demonstrated that even static magnetic islands can significantly modify linear drift-wave behavior by broadening the toroidal mode spectrum, coupling multiple mode numbers, and redistributing potential fluctuation structures in response to island-induced profile flattening. These results underscore the importance of incorporating island effects within linear gyrokinetic frameworks as a necessary step toward fully nonlinear analyses.

\noindent Experimental observations from the ADITYA-U tokamak provide strong validation for the magnetic island configurations considered in this work. As reported in \cite{Tanna2019}, Mirnov-coil phase analysis routinely identifies a coherent $(m,n)=(2,1)$ magnetic island at the $q=2$ resonant surface, with a measured width of a few centimeters and clear evidence of density and temperature flattening inside the island region. This flattening reflects profile relaxation, in which particle motion along the reconnected magnetic field lines repeatedly samples the island flux surfaces \cite{hornsby2011interaction}, reducing local gradients and producing nearly uniform profiles within the island region. These experimentally established characteristics—mode helicity, resonant location, island width, and the associated equilibrium profile modification—are incorporated into our perturbed equilibrium and reproduced self-consistently during the simulation relaxation phase. In addition to this dominant tearing mode, ADITYA-U measurements also reveal magnetic activity near the $q=3$ resonant surface, with Mirnov spectra indicating the presence of higher-order modes consistent with $(m,n)=(3,1)$ island \cite{patel2023role}. Owing to the higher safety factor and longer magnetic connection lengths, defined as the distance along a magnetic field line between successive poloidal transits, which scales approximately as $L_c\sim qR$\cite{Wesson2011} at the $q=3$ surface, such islands generate equilibrium perturbations that are more spatially extended but less localized than those at $q=2$, leading to broader and milder profile modifications. By modeling islands at both $q=2$ and $q=3$ with experimentally consistent topology and spatial scale, our simulations capture the distinct equilibrium modifications associated with each resonant surface and provide a realistic, experimentally grounded foundation for the ITG stability analysis that follows.

\noindent The role of magnetic islands in modulating plasma turbulence and transport is an active area of research and is central to optimizing tokamak performance. However, their influence on the linear ion-temperature-gradient (ITG) instability remains insufficiently explored. A Gyrokinetic theory predicts that small magnetic islands can stabilize ITG modes and alter their spatial structure in a sheared slab geometry \cite{wilson2009influence}. These predictions motivate a systematic investigation in realistic tokamak configurations. The ADITYA-U tokamak, with well-diagnosed plasma conditions and accessible parameter regimes, therefore provides a suitable platform for examining the interaction between magnetic islands and ITG-driven fluctuations \cite{Tanna2019}.

\noindent Advanced diagnostics, such as Thomson scattering, Langmuir probes, and bolometric measurements, further enable detailed characterization of equilibrium and fluctuation properties in the presence of magnetic islands. Building on such experimental capabilities, recent work on KSTAR \cite{Wei2025KSTAR} has used global gyrokinetic simulations to explore the nonlinear interaction between magnetic islands and ITG turbulence, revealing enhanced particle and heat transport arising from turbulence–zonal-flow–island coupling.

\noindent The present study seeks to bridge theoretical understanding and experimental observations by examining how static magnetic islands influence the linear stability of ITG modes in tokamak plasmas. By analyzing the growth rates of different toroidal modes and their convergence at large island widths, we assess whether island-induced equilibrium modifications can stabilize dominant ITG modes. Insights from this work provide a physics-based foundation for future studies of island control and confinement optimization, including techniques such as electron cyclotron current drive (ECCD)\cite{LaHaye2006} and resonant magnetic perturbations (RMPs) .

\noindent We carry out this study using the Global Gyrokinetic Code in Cylindrical Coordinates (G2C3) \cite{alageshan2024neural}, a particle-in-cell (PIC) code formulated in cylindrical geometry. In G2C3, field-line-aligned gather–scatter operations are performed using a neural-network-assisted projection scheme, avoiding coordinate transformations and associated numerical singularities near the island boundary.

\noindent In this present work, G2C3 is employed to perform linear gyrokinetic simulations of ITG modes in the presence of static magnetic islands. The objective is to isolate the linear response of drift-wave instabilities to island-induced equilibrium profile modifications in a realistic ADITYA-U configuration. This study represents a first step toward future nonlinear investigations, where self-consistent island evolution, zonal flows, and turbulent transport will be incorporated.

\noindent This paper commences with an introduction in Sec.~\ref{sec:introduction}. Sec.~\ref{sec:gyrokinetic_framework} details the fundamental gyrokinetic simulation framework employed.Sec.~\ref{sec:augmenting_island} then describes the specific methodology for modelling the Magnetic Island. Following this, Sec.~\ref {sec:simulation_itg_island} outlines the two-phase simulation of ITG in the Presence of a Magnetic Island. The core findings, including the impact on ITG mode structures and growth rates, are presented in Sec.~\ref{sec:results}. Finally, the conclusions are in Sec.~\ref{sec:conclusion}.

\section{Gyrokinetic simulation with adiabatic electrons}\label{sec:gyrokinetic_framework}

\noindent This paper focuses on electrostatic collisionless processes with adiabatic electrons to study linear ion temperature gradient (ITG) modes. Appropriately, the dynamics of the ions are described by the Vlasov-Maxwell equations, which are given by,
\begin{eqnarray}
    \frac{\partial f_i}{\partial t} &+& \dot{\mathbf{x}} \cdot \nabla f_i + 
          \dot{\mathbf{v}} \cdot \frac{\partial f_i}{\partial \mathbf{v}} = 0  \; ;
    \;\;\;\;\;\; \nabla \cdot \mathbf{E} = 4\pi \rho
\end{eqnarray}
where $\mathbf{v} = \dot{\mathbf{x}}$, and $f_i(\mathbf{x},\mathbf{v})$ is the ion
distribution function in the 6D phase space; $\mathbf{E}$ is the electric
field and $\rho$ is the corresponding charge density. The above set of equations is
solved self-consistently using 
\begin{eqnarray}
    \dot{\mathbf{v}} = \left( \frac{q}{m}\mathbf{E} + \frac{q}{mc} \mathbf{v} \times
            \mathbf{B} \right) \;\; ; \;\;\;\;\;  \rho = e \int \; d^3v \; (f_i-f_e),
\end{eqnarray}
where $\mathbf{B}$ is the applied external magnetic field and $f_e$ corresponds
to the Maxwellian distribution of the adiabatic electron.

\noindent We study microturbulence phenomena with much larger time scales
than the gyro-motion time scales of ions and electrons, which enables us to integrate
out the gyro-motion. This reduction leads to an effective, computationally efficient, 5D
gyro-kinetic phase space, $(R,\zeta,Z,v_{||},\mu)$, and the
corresponding evolution equation is expressed as~\cite{Singh2023,De2019Kinetic,Singh_2024}
\begin{eqnarray}
   \frac{d}{dt} f_i = \frac{\partial f_i}{\partial t} + \dot{\mathbf{x}} \cdot \nabla f_i + \dot{v}_{\parallel} \frac{\partial f_i}{\partial v_{\parallel}} = 0,
   \label{eq:GVM}
\end{eqnarray}
 where $f_i$ is now the guiding center distribution function.

\noindent We use G2C3~\cite{alageshan2024neural} to perform the gyrokinetic simulation in cylindrical
coordinates. As a gyro-kinetic PIC simulation,
instead of evolving the 5D partial differential equation (PDE) for $f_i$, we evolve the $N_i$ \enquote{marker particles}, thus reducing the problem to $5N_i$ ordinary differential
equations (ODE) for the guiding centers: 
\cite{alageshan2024neural,Singh_2022,Singh_2024,Singh_2024a}
\begin{eqnarray}
   \dot{\mathbf{X}} = v_{\parallel} \frac{\mathbf{B}}{B_\parallel^*} + \mathbf{v}_E 
             + \mathbf{v}_d, \;\; ; \;\;\;\;\;
   \dot{v}_{\parallel} = - \frac{1}{m_i} \frac{\mathbf{B}^*}{B_\parallel^*} \cdot \left( \mu \nabla B + Z_i \nabla \phi \right),
   \label{ParticleDynamics1}
\end{eqnarray}

\noindent and $\mathbf{B}^* = \mathbf{B} + B v_\parallel(\nabla\times \hat{b})/\omega_{ci}$ is the equilibrium magnetic field at the guiding center position, where $\omega_{ci}=q_{i}B/m_{i}c$, $B_\parallel^*=\hat b\cdot\mathbf{B}^*$, $\mathbf{v}_E$ is the ${E}\times{B}$ drift velocity, $q_i=Z_{i}e$(ion charge state($Z_i$) times the elementary charge(e)) and $\mathbf{v}_d$ is the magnetic drift velocity due to curvature and gradient in magnetic field, which are given by
\begin{eqnarray}
   \mathbf{v}_E = \frac{\hat{b}\times \nabla \phi}{B} \;\;\;\;\;\; \text{and} \;\;\;\;\;\; 
   \mathbf{v}_d = \mathbf{v}_c + \mathbf{v}_g = 
   \frac{v_\parallel^2}{\omega_{ci}} \nabla\times \hat{b} + \frac{\mu}{m_i \omega_{ci}} \hat{b}\times \nabla B.
   \label{drift_terms}
\end{eqnarray}

\noindent Now, if the equilibrium current is suppressed, then
\begin{eqnarray}
 \mathbf{v}_d &=& \left( \frac{m_i \; v_\parallel^2 + \mu \; B}{m_i \; \omega_{ci}} \right) \left( \frac{\hat{b}\times \nabla B}{B} \right)
\end{eqnarray}

\noindent The weights of the marker particles are then projected to the poloidal grids along the field lines, using
the neural network projection operator in G2C3~\cite{alageshan2024neural}. If $\delta n_i$ is the number density of 
projected ions on the poloidal plane, then the corresponding potential $\phi$ is computed using the
finite-element-based gyrokinetic-Poisson solver, employing adiabatic electron model, given by~\cite{alageshan2024neural}
\begin{eqnarray}
    \left(1 - \frac{1+\tau}{\tau} \nabla^2_\perp\right) \phi = Z_i\delta n_i,
\end{eqnarray}
where, $\tau=T_e/T_i$. The gyro-average of the potential is calculated using the 4-point method~\cite{lee1987gyrokinetic}. The gradients of the gyro-averaged potential provide the force term in Eq.~\ref{ParticleDynamics1} which advances the particles. The updated particle
positions are then used to compute the new potential, completing the PIC cycle. Energy-conserving boundary conditions are employed to return particles that leave the simulation domain. For all fluctuating quantities, fixed boundary conditions are applied at both radial boundaries of the domain.

\noindent In this paper, we perform linear simulations of the ITG mode, analyzing the evolution of instabilities under the assumption that perturbations are small and evolve exponentially without nonlinear coupling. In this linear framework, particle trajectories are determined entirely by the equilibrium magnetic field; particles follow unperturbed field lines shaped by the background toroidal and poloidal magnetic field structures of the tokamak. This setup establishes a reference case for the ITG mode dynamics in the absence of magnetic perturbations. To study the influence of magnetic islands on ITG-driven instabilities, we introduce a localized magnetic perturbation by modifying the equilibrium flux functions, as explained in Sec.~\ref{sec:augmenting_island}

\section{Augmenting the magnetic island} 
\label{sec:augmenting_island}

\noindent We begin with the experimentally reconstructed flux functions from ADITYA-U and introduce a magnetic island by adding a controlled helical perturbation to them. This perturbation is implemented through the analytic expressions for the radial, poloidal, and toroidal components of the island magnetic field, which modify the poloidal flux and produce the desired (m,n) structure at the specified resonant surface\cite{Jiang2014}, as
\begin{equation}
    \mathbf{B} = \mathbf{B}_0 + \delta\mathbf{B}_I \;\;\;\;\; \text{and}  \;\;\;\;\;\;
   \mathbf{B}^* = \mathbf{B_0^*} + \frac{B_0 v_\parallel}{\omega_{ci}} (\nabla\times \hat{\mathbf{b}}_0) + \delta\mathbf{B}_I 
\label{ParticleDynamics}
\end{equation}
For cylindrical coordinates with contravariant basis vectors $e^{R}=\nabla R$, $e^{\zeta}=\nabla \zeta$, $e^{Z}=\nabla Z$ and $e^i=g^{ij}e_{j}$, the metric tensor components are 
$g_{RR} = 1, \quad g_{\zeta\zeta} = R^2, \quad g_{ZZ} = 1$ and orthogonality implies that all off-diagonal terms are zero. The equilibrium magnetic field is therefore written as
\begin{equation}
\mathbf{{B_0}} = {B}^R \: \mathbf{e}_R + {B}^Z \: \mathbf{e}_Z + {B}^\zeta \: \mathbf{e}_\zeta 
= {B}_R \: \mathbf{e}^R + {B}_Z \: \mathbf{e}^Z + (R{B}_\zeta) \: \mathbf{e}^\zeta
\end{equation}

\noindent which in terms of the poloidal flux-function $\psi(R,Z)$ - labeling the magnetic
surfaces for both closed and open ﬁeld lines, and the poloidal current function $F(\psi)$ is given by \cite{De2019Kinetic} 
\begin{equation}
    \mathbf{B_0} = \nabla \psi(R, Z) \times \nabla \zeta + \frac{F(\psi)}{R} \hat{\zeta}
\end{equation}
which provides the components of the magnetic field in cylindrical coordinates as,
\begin{equation}
    {B_R} = -\frac{1}{R}\frac{\partial \psi}{\partial Z}, \quad {B_Z} = \frac{1}{R}\frac{\partial \psi}{\partial R}, \quad {B_\zeta} = \frac{F(\psi)}{R}
\end{equation}

\noindent  Now, the perturbation is modeled using the parameter $\alpha(R,Z,\zeta)$ (which have dimension of length and $\alpha\ll1$ ), with $\mathbf{A}_I = \alpha \mathbf{B}_0$,
  such that $\delta \mathbf{B}_I = \nabla\times\alpha \mathbf{B}_0$. In cylindrical coordinates, we get,
\begin{eqnarray}
   \delta B_{RI} &:=& \delta \mathbf{B}_I \cdot \nabla R =-\frac{R}{\mathcal{J}}\left(\alpha\frac{\partial B_{\zeta}}{\partial Z}+B_{\zeta}\frac{\partial \alpha}{\partial Z}\right)
 \\
   \delta B_{ZI} &:=& \delta \mathbf{B}_I \cdot \nabla Z =\frac{1}{\mathcal{J}}\left(\alpha R\frac{\partial B_{\zeta}}{\partial R}+R B_{\zeta}\frac{\partial \alpha}{\partial R}+\alpha B_{\zeta}\right)
   \\
   \delta B_{\zeta I} &:=& \delta \mathbf{B}_I \cdot \nabla \zeta =-\frac{1}{\mathcal{J}}\left(\alpha\frac{\partial B_{Z}}{\partial R}+B_{Z}\frac{\partial \alpha}{\partial R}\right) +\frac{1}{\mathcal{J}}\left(\alpha\frac{\partial B_{R}}{\partial Z}
   +B_{R}\frac{\partial \alpha}{\partial Z}\right)
\end{eqnarray}
where $\mathcal{J}^{-1}=\nabla R.(\nabla\zeta\times\nabla Z)$ is the Jacobian for this system

\noindent Similarly, we can write the velocity components and $\dot{v}_{\parallel}$ in cylindrical coordinates using Eq. \ref{ParticleDynamics} in the presence of the perturbed magnetic field, and for an axisymmetric system, the components are as follows:

\begin{eqnarray}
   \dot R &=& {v_{||}}\frac{{\tilde{{B_R}}}}{{{B_{||}^*}}} + \frac{c}{{{B_{||}^*}}}\left( {\frac{{\tilde{{B_\zeta }}}}{{B}}\frac{{\partial \phi }}{{\partial Z}} - \frac{{\tilde{{B_Z}}}}{{R{B}}}\frac{{\partial \phi }}{{\partial \zeta }}} \right)  
 - \frac{{{B}v_{||}^2}}{{{B_{||}^*}{\omega _{ci}}}}\frac{1}{\mathcal{J}}\frac{\partial }{{\partial Z}}\left( {\frac{{R\tilde{{B_\zeta }}}}{{B}}} \right) + \frac{\mu }{{m_i{\omega _{ci}}}}\frac{{\tilde{{B_\zeta }}}}{{{B}}}\frac{{\partial {B}}}{{\partial Z}}
 \label{dens2} \\
  \dot Z &=& {v_{||}}\frac{{\tilde{{B_Z}}}}{{{B_{||}^*}}} - \frac{c}{{{B_{||}^*}}}\left( {\frac{{\tilde{{B_\zeta }}}}{{B}}\frac{{\partial \phi }}{{\partial R}} - \frac{{\tilde{{B_R}}}}{{R{B}}}\frac{{\partial \phi }}{{\partial \zeta }}} \right)  + \frac{{{B}v_{||}^2}}{{{B_{||}^*}{\omega _{ci}}}}\frac{1}{\mathcal{J}}\frac{\partial }{{\partial R}}\left( {\frac{{R\tilde{{B_\zeta}}}}{{B}}} \right) - \frac{\mu }{{m_i{\omega _{ci}}}}\left( {\frac{{\tilde{{B_\zeta }}}}{{{B}}}\frac{{\partial {B}}}{{\partial R}}} \right) \\
  \dot\zeta &=& {v_{||}}\frac{{\tilde{{B_\zeta }}}}{{{B_{||}^*}}}\frac{1}{R} + \frac{c}{{{B_{||}^*}\mathcal{J}}}\left( {\frac{{\tilde{{B_Z}}}}{{B}}\frac{{\partial \phi }}{{\partial R}} - \frac{{\tilde{{B_R}}}}{{B}}\frac{{\partial \phi }}{{\partial Z}}} \right) 
 - \frac{\mu }{{m_i{\omega _{ci}}\mathcal{J}}}\left( {\frac{{\tilde{{B_R}}}}{{{B_{||}^*}}}\frac{{\partial {B}}}{{\partial Z}} - \frac{{\tilde{{B_Z}}}}{{{B_{||}^*}}}\frac{{\partial {B}}}{{\partial R}}} \right)\nonumber\\
    && + \frac{{{B}v_{||}^2}}{{{B_{||}^*}{\omega _{ci}}}}\frac{1}{\mathcal{J}}\left[ {\frac{\partial }{{\partial Z}}\left( {\frac{{\tilde{{B_R}}}}{{B}}} \right) - \frac{\partial }{{\partial R}}\left( {\frac{{\tilde{{B_Z}}}}{{B}}} \right)} \right] \\
    \dot v_{||} &=& - \frac{\mu }{m_i}\left[ {\frac{{\tilde{{B_R}}}}{{{B_{||}^*}}}\frac{{\partial {B}}}{{\partial R}} + \frac{{\tilde{{B_Z}}}}{{{B_{||}^*}}}\frac{{\partial {B}}}{{\partial Z}}} \right] 
     - \frac{{{Z_i}}}{m_i}\left[ {\frac{{\tilde{{B_R}}}}{{{B_{||}^*}}}\frac{{\partial \phi }}{{\partial R}} + \frac{{\tilde{{B_\zeta}}}}{{R{B_{||}^*}}}\frac{{\partial \phi }}{{\partial \zeta }} + \frac{{\tilde{{B_Z}}}}{{{B_{||}^*}}}\frac{{\partial \phi }}{{\partial Z}}} \right]\nonumber\\
    && - \frac{{\mu {v_{||}}}}{{m_i{\omega _{ci}}}}\frac{{B}}{{{B_{||}^*}}}\frac{1}{\mathcal{J}}\left[ {\frac{\partial }{{\partial R}}\left( {\frac{{R\tilde{{B_\zeta }}}}{{B}}} \right) - R\frac{\partial }{{\partial Z}}\left( {\frac{{\tilde{{B_\zeta }}}}{{B}}} \right)} \right]
     - \frac{{{v_{||}}{Z_i}}}{{m_i{\omega _{ci}}}}\frac{{B}}{{{B_{||}^*}}}\frac{1}{\mathcal{J}}\left[ {\frac{{\partial \phi }}{{\partial Z}}\frac{\partial }{{\partial R}}\left( {\frac{{R\tilde{{B_\zeta }}}}{{B}}} \right)} \right. \nonumber \\
    && \left.{+ \frac{{\partial \phi }}{{\partial \zeta }}\left( {\frac{\partial }{{\partial Z}}\left( {\frac{{\tilde{{B_Z}}}}{{B}}} \right) - \frac{\partial }{{\partial R}}\left( {\frac{{\tilde{{B_Z}}}}{{B}}} \right)} \right) } \right. 
     \left. {- R\frac{{\partial \phi }}{{\partial R}}\frac{\partial }{{\partial Z}}\left( {\frac{{\tilde{{B_\zeta }}}}{{B}}} \right)} \right].
\label{density evolution}
\end{eqnarray}
where $B=|{\mathbf{B}}|$, $\tilde{B_R} = B_R + \delta B_{RI}$, $\tilde{B_\zeta}=B_{\zeta} + \delta B_{\zeta I}$ and $\tilde{B_Z}=B_Z + \delta B_{ZI}$.


\noindent Now, we define the operator $ \mathcal{L} := \frac{d}{dt} = \frac{\partial }{\partial t} + \dot{\mathbf{x}} \cdot \nabla  + \dot{v}_{\parallel} \frac{\partial }{\partial v_{\parallel}}$, such that the collisionless gyrokinetic Vlasov equation is
$\mathcal{L}f=0$. The operator $\mathcal{L}$ is split into equilibrium and perturbation parts,  i.e.,
$\mathcal{L}=\mathcal{L}_0+\delta \mathcal{L}$, such that

\begin{eqnarray}
    \mathcal{L}_0= \frac{\partial}{\partial t} + \left( v_{\parallel} \frac{(\mathbf{B_0}+\delta\mathbf{B}_I)}{\mathbf{B_\parallel^*}}  + \mathbf{v}_d \right)\cdot \nabla - \frac{\mu}{m_i}\frac{(\mathbf{B_0^*}+\delta\mathbf{B}_I)}{\mathbf{B_0}} \cdot \nabla \mathbf{B_0} \frac{\partial}{\partial v_\parallel} \;\;,
    \;\;\;\;\text{and}
\end{eqnarray}

\begin{eqnarray}
     \delta \mathcal{L} &=&  
     \mathbf{v}_E \cdot \nabla 
     -  \frac{Z_i}{m_i}\frac{\mathbf{B}^* }{\mathbf{B_0}} \cdot \nabla \phi \frac{\partial}{\partial v_\parallel}
\end{eqnarray}

\noindent Let us split the distribution function similarly,i.e., $f=f_0+\delta f$. Using the fact that
the equilibrium part satisfies $\mathcal{L}_0f_0 = 0$ and subtracting it from $\mathcal{L}f = 0$, we get $\mathcal{L}\delta f 
= - \delta \mathcal{L} f_0$. For linear simulation, weight is defined as $w=\frac{\delta f}{f_0}$. Then
\begin{eqnarray}
    \mathcal{L}w = \frac{1}{f}\mathcal{L}\delta f = -\frac{1}{f}\delta \mathcal{L} f_0   
    = -(1-w)\left(\frac{1}{f_0}\delta \mathcal{L} f_0\right)
\end{eqnarray}
\begin{equation}
    \Rightarrow\;\;\; \mathcal{L}_0 w = -\frac{1}{f_0}\delta \mathcal{L} f_0
\end{equation}

So, the linear weight equation with island perturbation can be written as

\begin{eqnarray}
     \mathcal{L}_0w &=& -\left( v_{\parallel} \frac{\delta \mathbf{B}_I}{B_\parallel^*} + \mathbf{v}_E \right) \cdot \frac{\nabla f_0}{f_0} + \left( \frac{\mu}{m_i}\frac{\delta\mathbf{B}_I}{B_0} \cdot \nabla B_0  + \frac{Z_i}{m_i}\frac{\mathbf{B}^* }{B_0} \cdot \nabla \phi \right) \frac{1}{f_0}\frac{\partial f_0}{\partial v_\parallel} \nonumber \\
     &=& - \underbrace{\mathbf{v}_E \cdot \frac{\nabla f_0 |_{v_\perp}}{f_0}}_{\text{Drive term}} - \underbrace{\frac{Z_i}{T_i}\left( \hat b\cdot\nabla\phi \right) v_\parallel}_{\text{Parallel term}}
   - \underbrace{\frac{Z_i}{T_i} \left(\mathbf{v}_d \cdot\nabla\phi \right) }_{\text{Drift term}} \nonumber \\ 
   && \;\;\; - \underbrace{\left( v_{\parallel} \frac{\delta \mathbf{B}_I}{B_\parallel^*} \cdot \frac{\nabla f_0 |_{v_\perp}}{f_0} + \frac{Z_i}{T_i}v_\parallel \frac{\delta\mathbf{B}_I}{B_\parallel^*} \cdot \nabla \phi \right)}_{\text{Island perturbation source term}}
\end{eqnarray}
where $\nabla f_0 |_{v_\perp}=\left(\nabla+\frac{\mu\nabla B_0}{T_0}\right)f_0$. The influence of magnetic islands on drift terms is ignored, resulting in $B_0^* = B_\parallel^* = B_0$ \cite{Li2023}, and thus the island perturbation terms are derived accordingly
\begin{eqnarray}
   \left(v_{\parallel} \frac{\delta\mathbf{B}_I}{B_\parallel^*} \cdot \frac{\nabla f_0 |_{v_\perp}}{f_0}
   \right)
    &=& \frac{v_\parallel}{f_0} \left[ \frac{\delta B_{IR}}{B_\parallel^*}\left(\frac{1}{n_i}\frac{\partial n_i}{\partial R}+ 
       \frac{\Gamma}{T_i}\frac{\partial T_i}{\partial R}\right)+\frac{\delta B_{IZ}}{B_\parallel^*}\left(\frac{1}{n_i}\frac{\partial n_i}{\partial Z}+ \frac{\Gamma}{T_i}\frac{\partial T_i}{\partial Z}\right)\right], \nonumber \\
   \left(\frac{Z_i}{T_i}v_\parallel \frac{\delta \mathbf{B}_I}{B_\parallel^*} \cdot \nabla \phi 
   \right)
    &=& \frac{Z_i v_\parallel}{T_i B_\parallel^*}  \left(\delta B_{IR} \frac{\partial \phi}{\partial R} + \delta B_{IZ}  \frac{\partial \phi}{\partial Z} + \frac{\delta B_{IZ}}{R}  \frac{\partial \phi}{\partial \zeta} \right)
    \label{density2}
\end{eqnarray}
where $\Gamma = \left( \frac{2\mu \; B + m_i \; v_\parallel^2 }{2 T_i } - \frac{3}{2} \right)$.

Now, to model the island with a mode number $(m,n)$, we need to write the perturbation in
terms of the Boozer angle, $\theta_B$, which can be extracted from the field line structure. 
Let the field lines starting from the outer mid-plane be given by
\begin{eqnarray}
    \mathbf{\mathcal{F}}(s):=\mathbf{\mathcal{F}}(R(s),Z(s),\zeta(s)) = \mathbf{X}(R_0,0,0) + 
    \int_0^s ds' \;\: \mathbf{\hat{b}}(R(s'),Z(s'),\zeta(s'))
\end{eqnarray}
where $s$ is the arc-length parameter along the field line and $R_0$ specifies the flux surface.
Now, to find $\theta_B$ of any point $(R,Z,\zeta)$, we use $\mathbf{B}$ at this point and move
along the field lines in the $\hat{\mathbf{b}}$ direction until the flux line crosses the
$Z=0$ plane on the outer radial direction at $(R_+,0,\zeta_+)$. Similarly, if starting
from $(R,Z,\zeta)$ if we move in the $-\hat{\mathbf{b}}$ direction and find the corresponding
point $(R_-,0,\zeta_-)$, then 
\begin{eqnarray}
   \theta_B(R,Z,\zeta) = \frac{2\pi \; \zeta_-}{\left( \zeta_+ - \zeta_- \right)} \;,
\end{eqnarray}
and the corresponding safety factor for the flux surface is given by
\begin{eqnarray}
   q(R,Z,\zeta) = q(\psi) = \frac{2\pi}{\left( \zeta_+-\zeta_- \right)}
\end{eqnarray}

\begin{figure}[!ht]
    \centering
    {\includegraphics[width=0.485\textwidth]{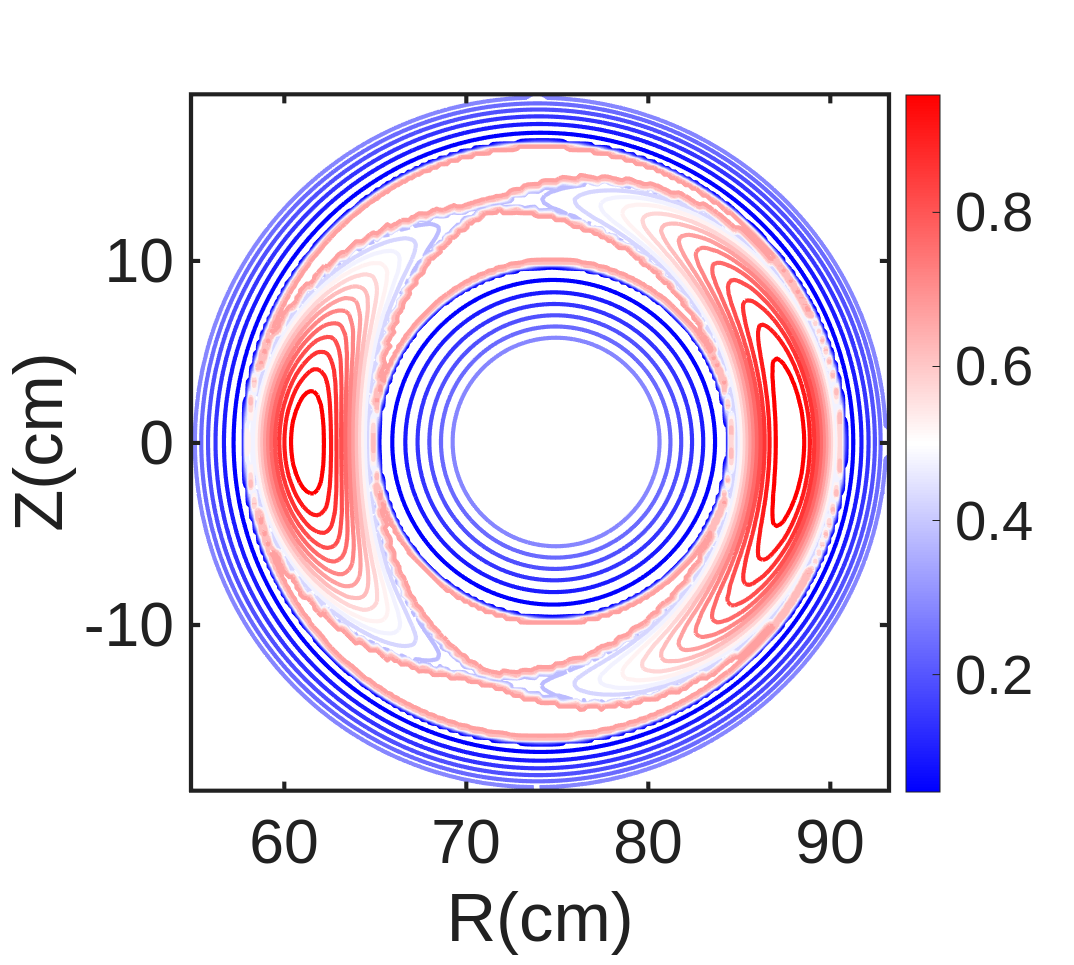}
    \includegraphics[width=0.48\textwidth]{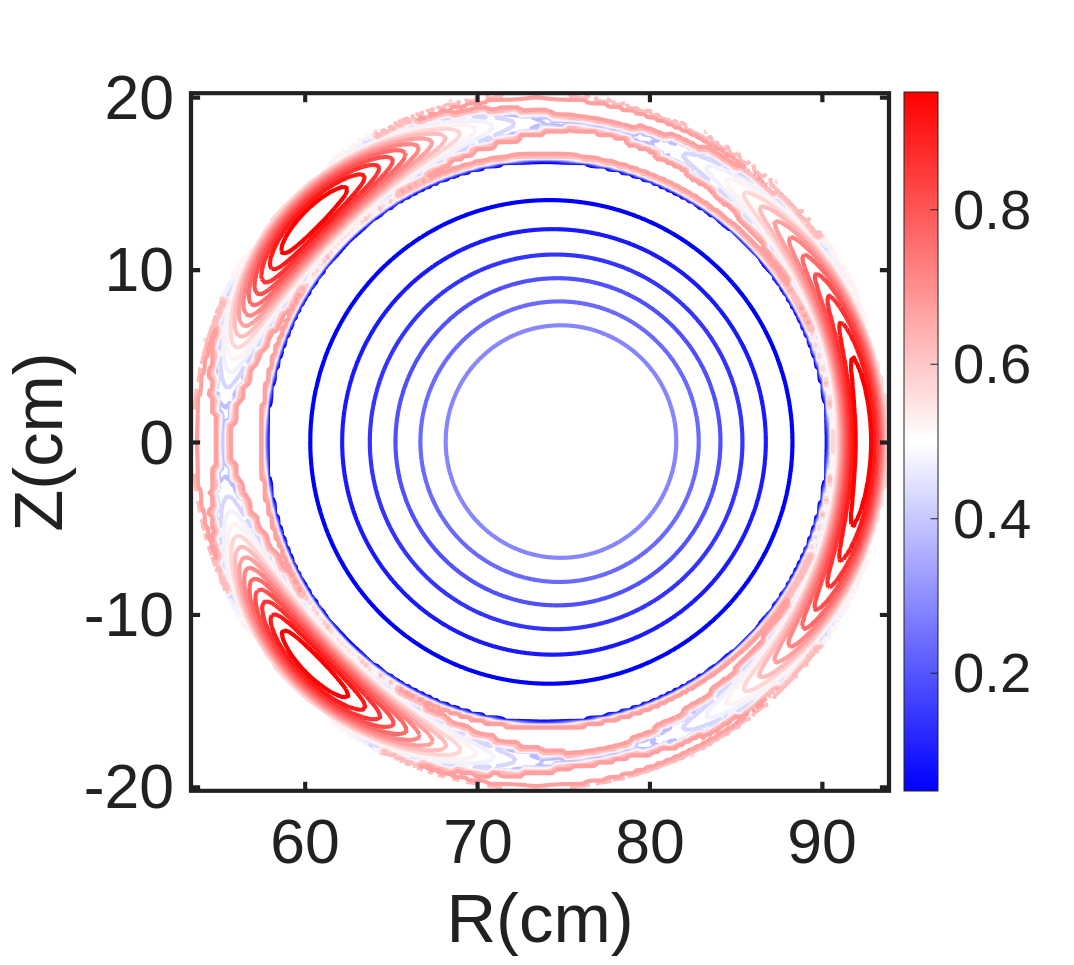}}
    \hspace{1cm} {(a)} \hspace{8cm} {(b)}
    \caption{(a) Perturbed helical magnetic flux surfaces $\tilde{\psi}_{he}$ showing the formation of a static $(m = 2,\; n = 1)$ magnetic island introduced at the $q = 2$ resonant surface in the ADITYA-U discharge (\#36628), plotted on the poloidal plane at $\zeta = 0$. 
    (b) Perturbed helical flux surfaces for a static $(m = 3,\; n = 1)$ magnetic island at the $q = 3$ surface for the discharge(\#32802).}
    \label{island}
\end{figure}

\begin{figure}[!ht]
    \centering
    {\includegraphics[width=0.485\textwidth]{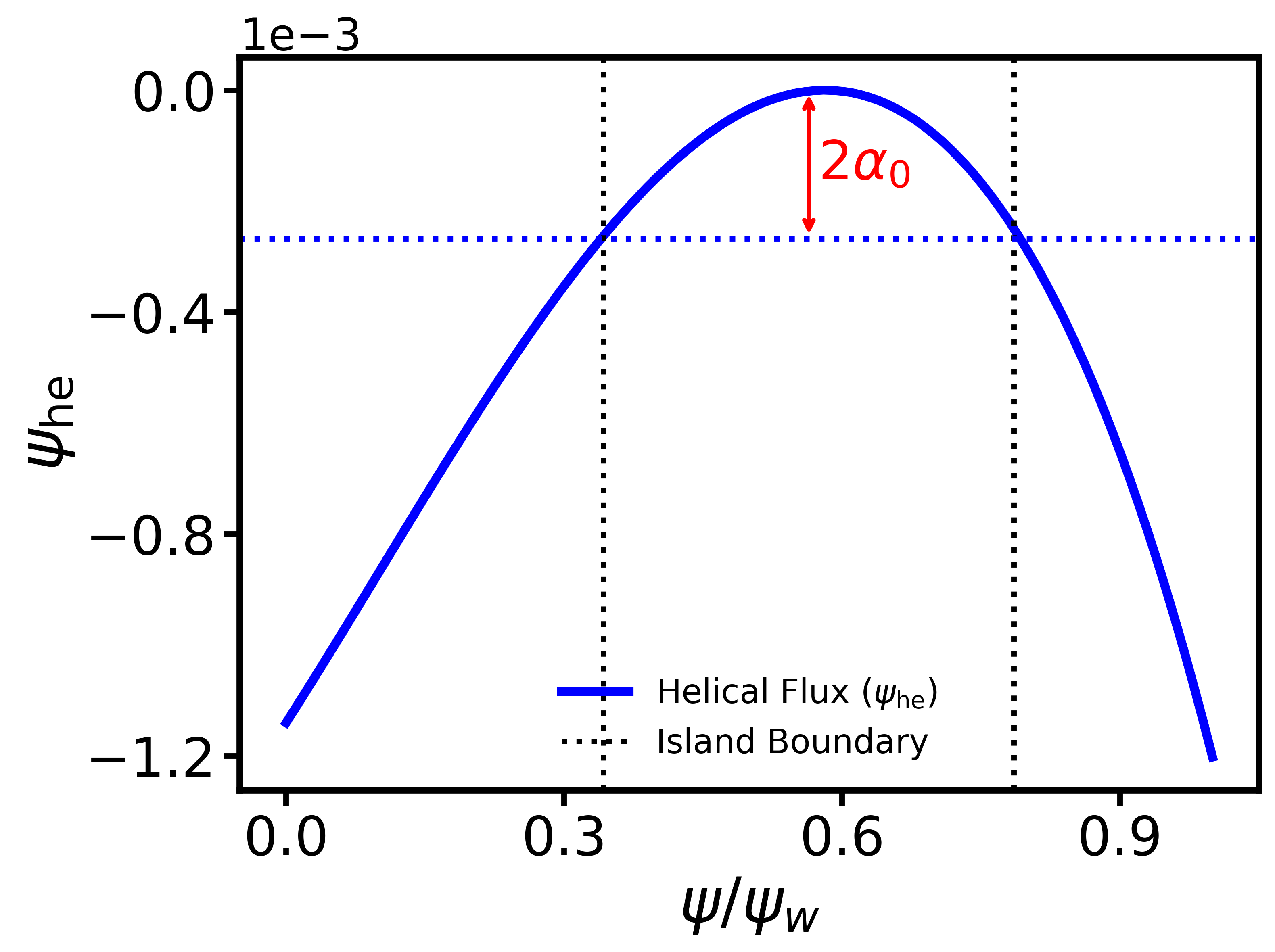}
    \includegraphics[width=0.48\textwidth]{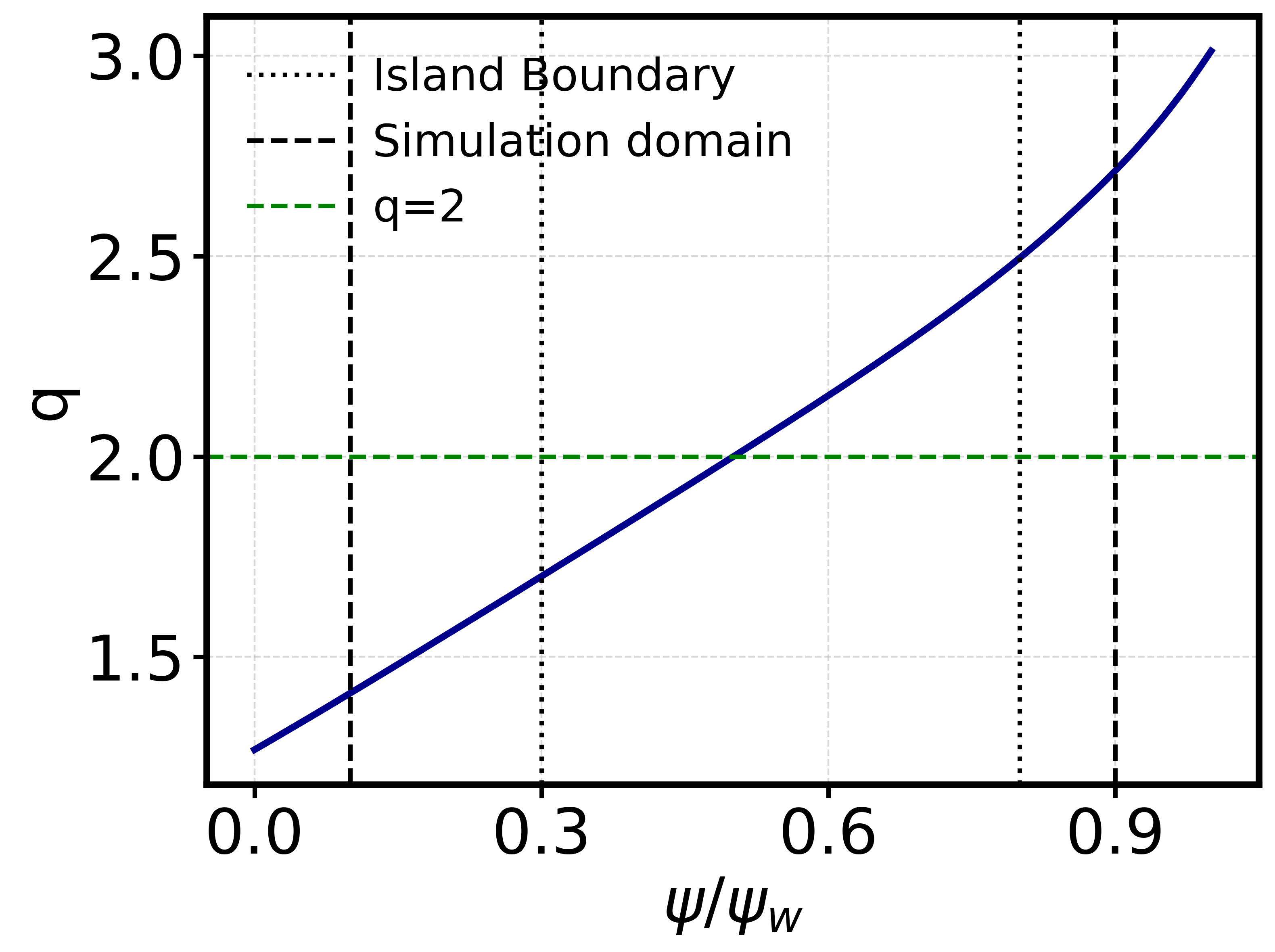}}
    \hspace{1cm} {(a)} \hspace{7cm} {(b)}
    \caption{(a) Schematic illustrating the magnetic-island width in flux-space using the perturbed helical flux function \(\tilde{\psi}_{he}\). (b) Safety-factor profile \(q(\psi)\) showing the resonant surface at which the island forms ( \(q=2\) for the \((m,n)=(2,1)\) case).
    }
    \label{helical}
\end{figure}

\noindent Finally, the $(m,n)$ island is specified by \(\alpha(\psi,\theta_B,\zeta) = \alpha_0(\psi) \cos(m \theta_B - n \zeta)\). Fig.~\ref{island}{(a)} shows the perturbed flux surfaces for $(2,1)$ island, and Fig.~\ref{island}{(b)} shows the perturbed flux surfaces for $(3,1)$ island. The perturbed helical flux function is given by \cite{Jiang2014},
\begin{eqnarray}
\tilde{\psi}_{he} =\psi_{he}-\alpha_0 \: g \; \cos m \xi = \left(\psi - \frac{\psi_t}{q_s}\right) - \alpha_0 \: g \; \cos m \xi ,
\end{eqnarray}
 where $\xi=\left(\theta_{B}-\frac{\zeta}{q_s}\right)$, $\psi_{he}=\left(\psi - \frac{\psi_t}{q_s}\right)$ is the equilibrium helical flux, $q_s$ is the safety factor of the resonant surface, $\psi_t$ is a toroidal flux function, and $g=\frac{F(\psi)}{2\pi}$ is the poloidal current function appearing in the Boozer representation of the magnetic field. The island, introduced through a perturbed vector potential 
 \begin{eqnarray}
 A_{I} = \alpha\mathbf{B_{0}}
 \end{eqnarray}
where the $\alpha(=\alpha_{0} \cos(m \theta_{B} - n\zeta))$ corresponds to the amplitude of the perturbation and has the units of $\psi_{he}$. In this simulation, we consider a single ion (Hydrogen) and assume the equilibrium ion density $n_{i0}$ to be identical to the electron density $n_{e0}$. Considering the $\psi_{he}$ of resonant surface has the highest value [Fig.~\ref{helical}{(a)}], the inner and outer walls of the magnetic islands corresponds to the flux surface which has an $\psi_{he}$ value of \enquote{$\psi_{he}$(resonant surface)-2$\alpha_0$} and from these inner and outer walls we can compute the width of island which is basically the distance between them. To simulate the $m=2$, $n=1$ magnetic island relevant to the ADITYA-U tokamak, we select the $q(=m/n)=2$ surface as the resonant flux surface [Fig.~\ref{helical}{(b)}]. In the ADITYA-U tokamak, magnetic islands are identified using an array of Mirnov coils placed at different poloidal and toroidal positions on the vacuum vessel. These coils detect the magnetic perturbations associated with rotating tearing modes, from which the island width has been experimentally estimated to be about 6 cm \cite{Tanna2018}. The corresponding poloidal (m) and toroidal (n) mode numbers are obtained by analyzing the phase differences of the Mirnov signals\cite{Tanna2019}. Each coil senses the same rotating perturbation with a phase shift determined by its spatial location. The poloidal phase variation between vertically separated coils gives the poloidal mode number m, while the toroidal phase variation between toroidally separated coils yields the toroidal mode number n. The relationship between phase and coil position is expressed as $\Delta\phi=m\Delta\theta+n\Delta\zeta$. To accurately model this in our simulation, the 6 cm physical width is first translated into its equivalent width in flux-space $(\psi-space)$.

\begin{figure}[!ht]
    \centering
    {\includegraphics[width=0.42\textwidth]{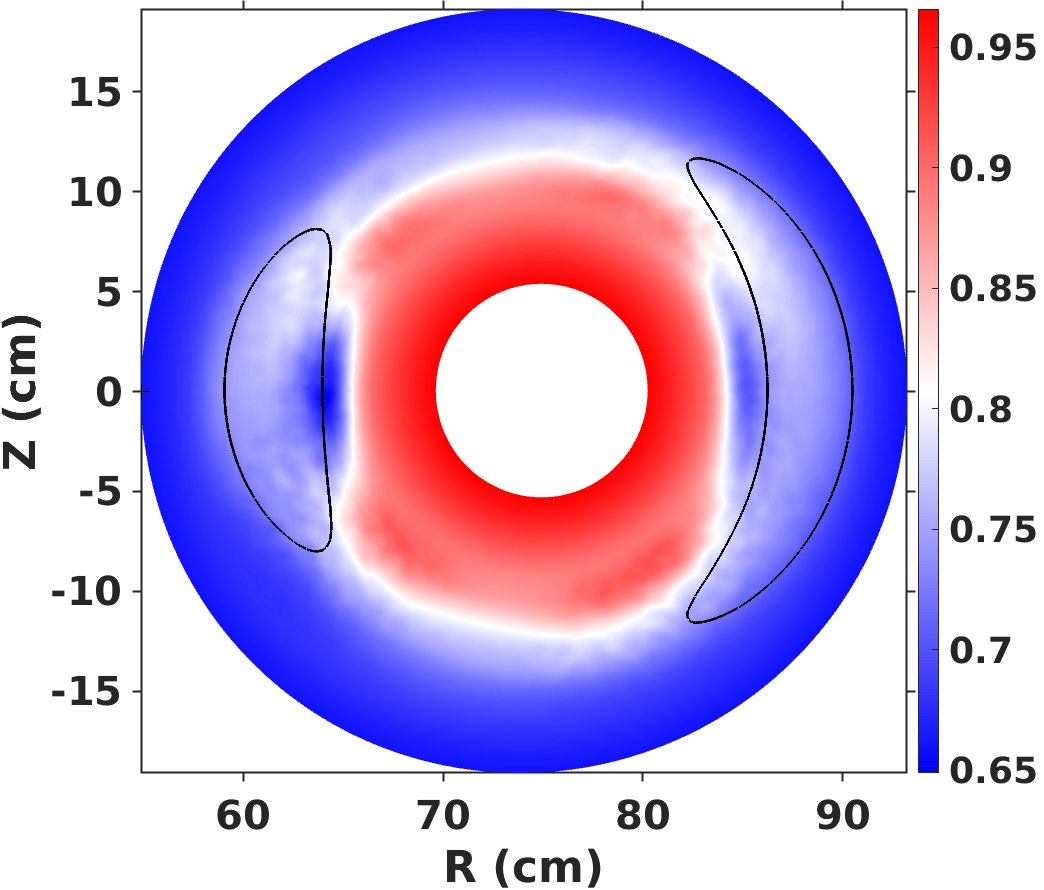}\hspace{0.5cm}
         \includegraphics[width=0.48\textwidth]{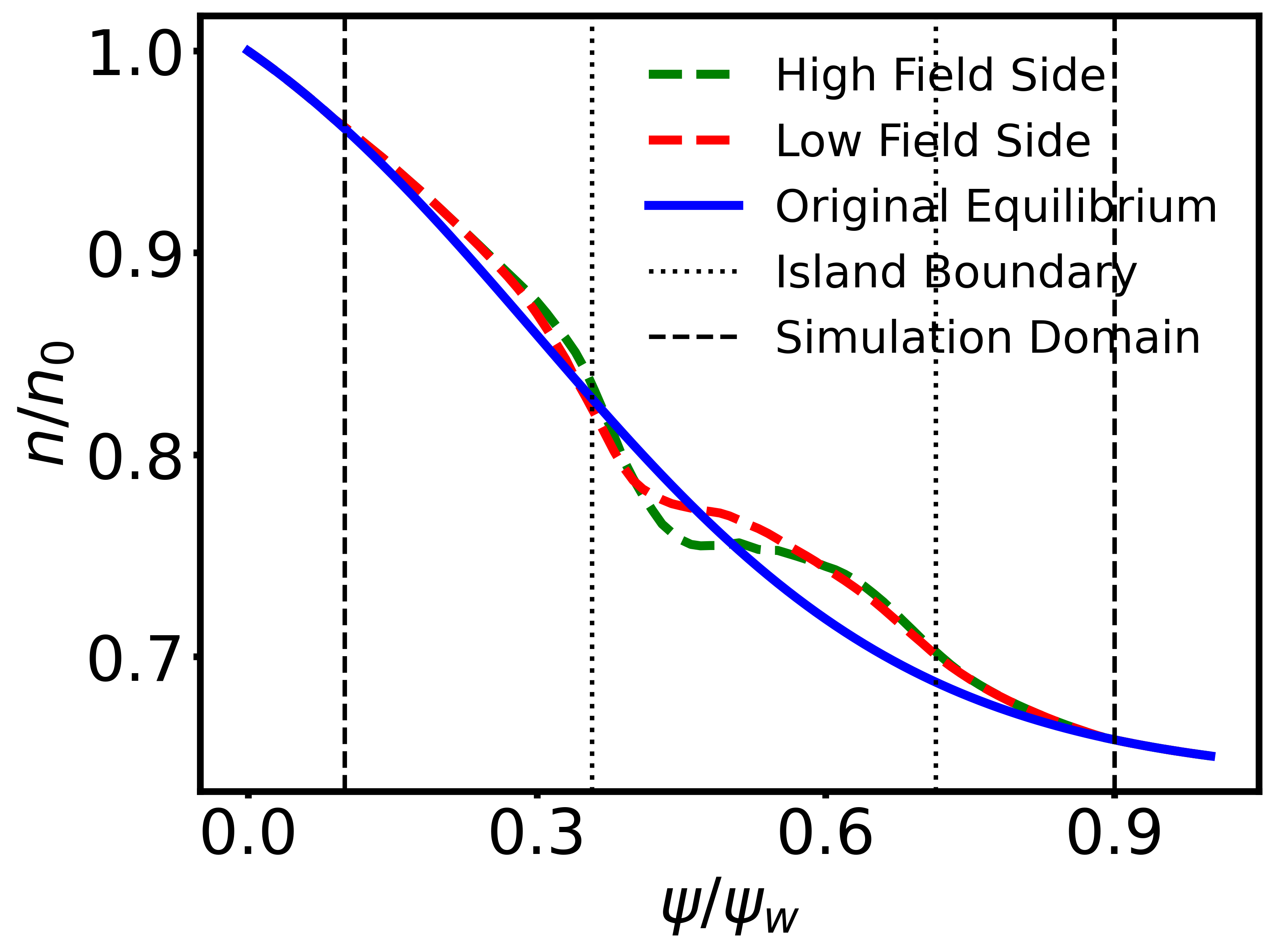}}
    \hspace{1cm} {(a)} \hspace{7cm} {(b)}
    \caption{(a) Poloidal cross-section showing the perturbed ion density from particle dynamics in the presence of a static $(m=2)$ magnetic island. Black contours represent the island boundaries. Density flattening is visible within the island region due to mirror-like trapping effects. 
     (b) Radial ion density profiles at $\theta = 0$ (low-field side) and $\theta = \pi$ (high-field side), compared with the unperturbed equilibrium profile. Flattening is more prominent on the high-field side, indicating mirror-induced confinement for island width $w = 0.29a$.}
    \label{flatten}
\end{figure}

\begin{figure}[!ht]
    \centering
    {\includegraphics[width=0.485\textwidth]{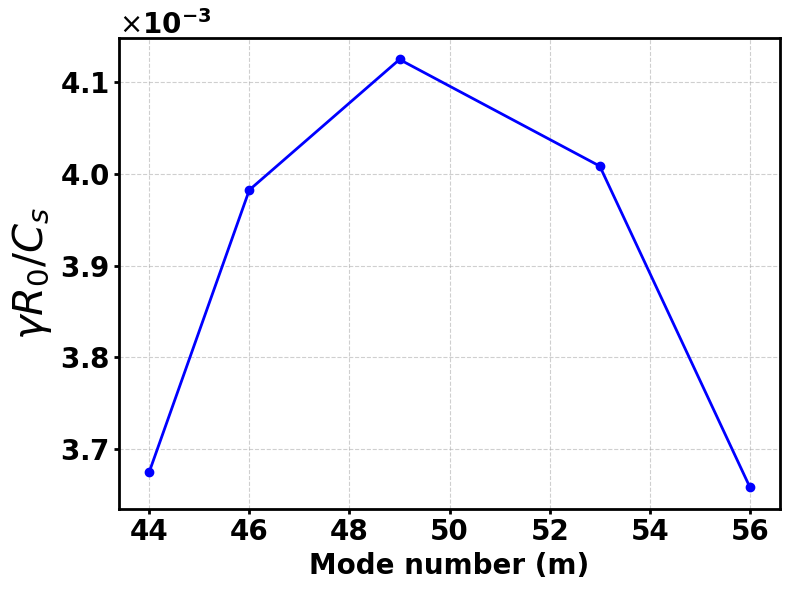}
    \includegraphics[width=0.48\textwidth]{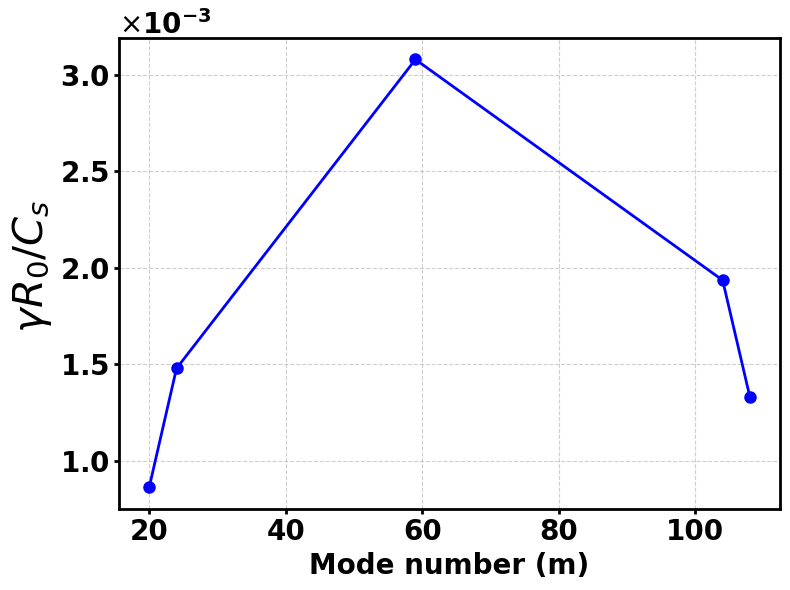}}
    \hspace{1cm} {(a)} \hspace{8cm} {(b)}
    \caption{Linear spectrum of the ion temperature gradient (ITG) mode obtained from global gyrokinetic simulations using the G2C3 code. (a) Spectrum for the equilibrium with gradients localized near the 
q=2 surface. (b) Corresponding spectrum for the equilibrium with gradients localized near the q=3 surface.}
    \label{dispersion}
\end{figure}

\section{Simulation of ITG in the Presence of Magnetic Island}\label{sec:simulation_itg_island}

\noindent
To investigate the influence of static magnetic islands on ion temperature gradient (ITG) instabilities, we follow a two-phase simulation strategy similar to that of Jiang et al.~\cite{Jiang2014}, implemented using the global gyrokinetic PIC code G2C3. In the first phase, a static $(m=2, n=1)$ magnetic island structure is introduced at the q=2 surface by perturbing the equilibrium magnetic flux as given in Eq.~\ref{ParticleDynamics} and the subsequent evolution of the density is obtained by running the particle dynamics Eq.~\ref {dens2}-~\ref{density2}, without the electric field. The plasma then relaxes without collision into a modified equilibrium characterized by a flattened ion density profile within the island region. During the relaxation phase, the density evolution is governed implicitly by the gyrokinetic equation for the distribution function. The resulting density flattening inside the island emerges self-consistently as an equilibrium feature due to the particle motion along the reconnected magnetic field lines in the modified magnetic topology. This flattening arises from parallel streaming of particles along the reconnected magnetic field lines. Because the mobility of charged particles is much higher along field lines than across them, any initial density or temperature gradients inside the island are rapidly equalized as particles stream freely along the closed helical field lines, effectively homogenizing the plasma within the island boundary.

\noindent The magnetic field strength is weaker at the island center and stronger near the island boundary, forming a magnetic mirror configuration. Owing to the conservation of magnetic moment ($d\mu/dt=0$), particles slow down as they approach regions of stronger magnetic field near the island boundary. If their parallel kinetic energy is insufficient to overcome the local mirror potential, they are reflected and become trapped in bounce orbits between the mirror points. This bounce-trapping mechanism produces localized confinement and enhances the homogenization of density within the island, particularly evident on the high-field side at $\theta = \pi$, as shown in Fig.~\ref{flatten}(b). These results are consistent with earlier gyrokinetic simulations using GTC for KSTAR~\cite{Wei2025KSTAR} and analytical equilibrium studies~\cite{Jiang2014}, both of which report profile flattening inside magnetic islands due to parallel streaming and mirror-trapped particle dynamics.

\noindent To further examine the linear characteristics of the ion temperature gradient (ITG) mode in the presence of magnetic islands, a detailed spectral analysis was performed using the G2C3 global gyrokinetic code. Fig.\ref{dispersion} presents the linear ITG mode spectra for two equilibria in which the background gradients are localized near the q=2 and q=3 resonant surfaces, corresponding respectively to $(m,n)=(2,1)$ and $(3,1)$ island configurations. These results represent the linear phase of the simulation, in which a static magnetic island is introduced by perturbing the equilibrium field, and the plasma response is evaluated without any nonlinear mode coupling. For the (2,1) island located at the q=2 surface, the ITG growth rate is $\sim 0.4$. The $(3,1)$ island at $q=3$ surface exhibits a similar spectral structure but with a slightly reduced growth rate of about 0.36. This stabilization can be understood in linear ITG theory, where instability arises from the competition between the density-gradient drive and the parallel dynamics of ions along magnetic field lines. At higher-q surfaces, the connection length increases approximately as $L_{c}\sim qR$ \cite{Wesson2011}, reducing the wavenumber $k\sim1/L_{c}$ and allowing particles moving rapidly along field lines to average out perturbations more effectively. As a result, the density and potential perturbations along the field line become less correlated, diminishing the ability of the curvature drift to extract free energy from the ion temperature gradient. In addition, the curvature drive itself $(\omega_{D}\propto 1/R)$ weakens at larger radius. Together, these geometric effects result in a smaller effective ITG drive and, consequently, a lower linear growth rate.
These linear-mode spectra serve as a reference for assessing how the presence of magnetic islands affects the ITG mode structure and stability, as discussed in the following section.

\noindent In the second phase, the flattened density profile obtained after island relaxation is used as the new equilibrium background for linear, electrostatic gyrokinetic simulations with adiabatic electrons, in which the equilibrium magnetic field is taken without the explicit island perturbation. This setup isolates the impact of island-induced equilibrium modification on ITG behavior. Small-amplitude electrostatic perturbations are evolved about this modified equilibrium, and their linear growth is monitored to determine the corresponding ITG growth rates and spatial structure of the ITG mode. Particular attention is given to the dependence of toroidal ITG mode growth rates on the island width, particularly whether larger islands lead to convergence of growth rates across different toroidal mode numbers.

\begin{figure*}
    \centering
    {\includegraphics[scale=0.30]{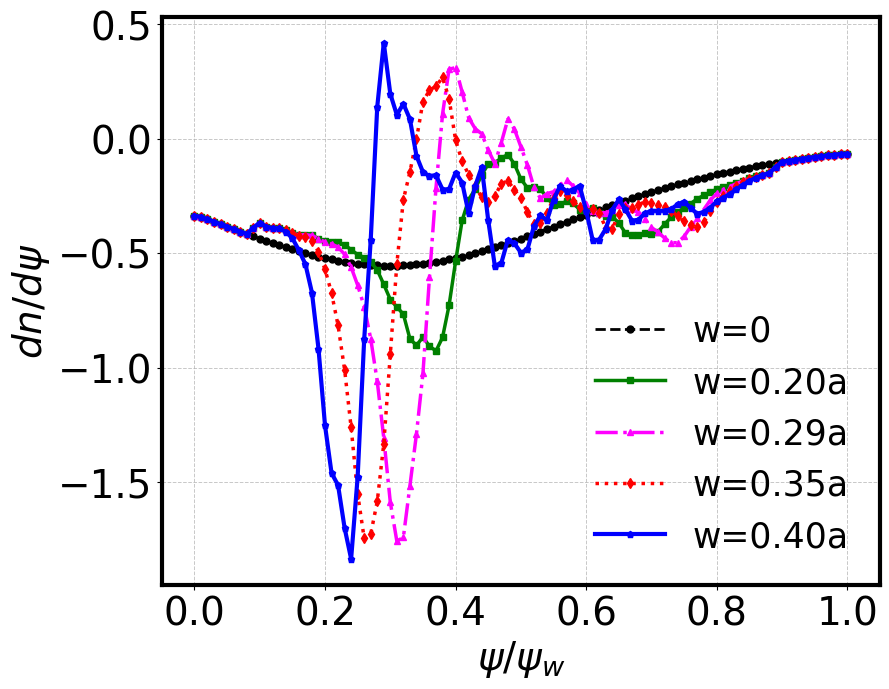}\hspace{1cm}
    \includegraphics[scale=0.35]{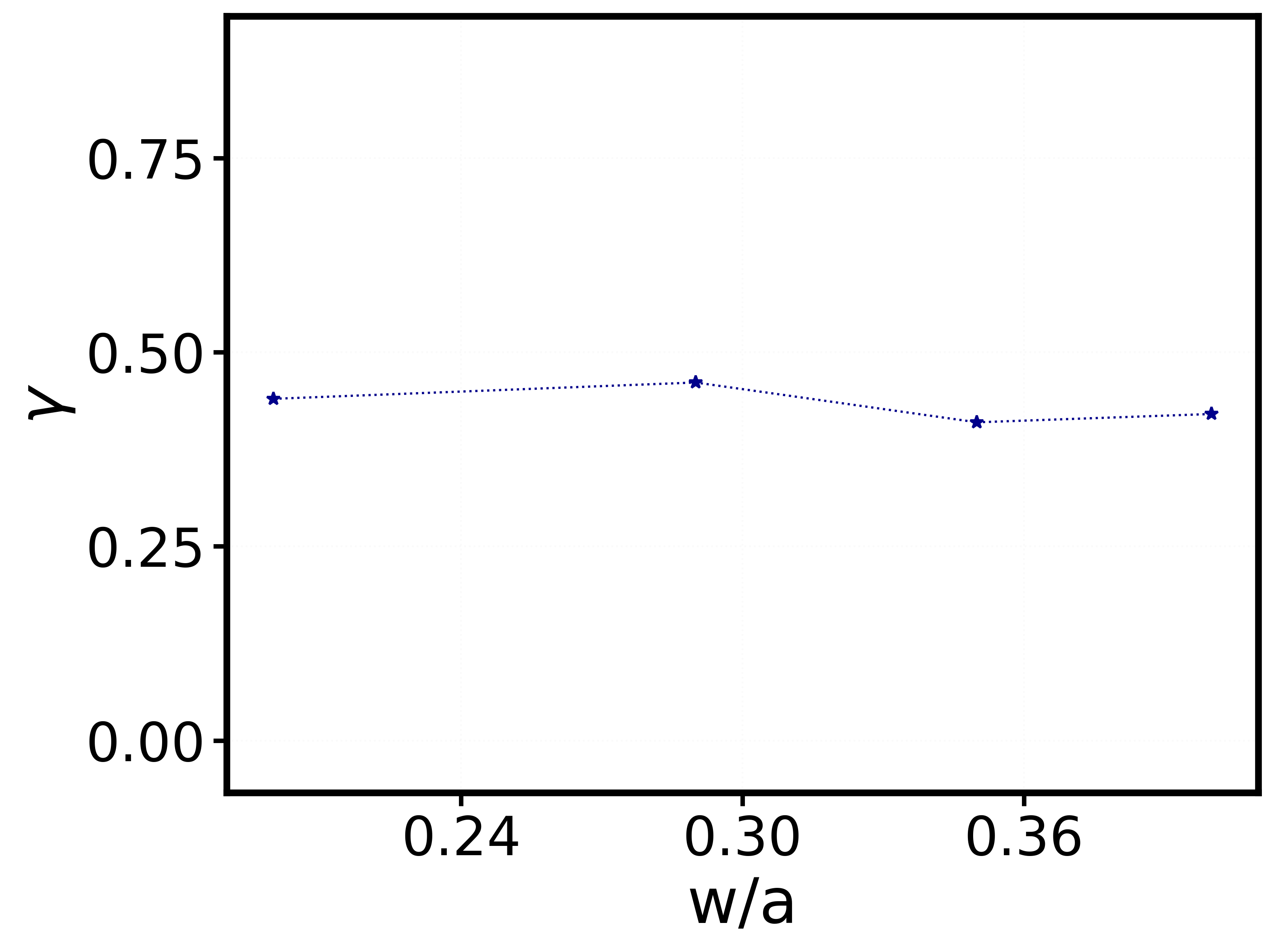}}\\
    \hspace{1cm} \hypertarget{density_grad}{(a)}  \hspace{8cm} \hypertarget{growth rate}{(b)}
    \caption{(a) Radial profile of the ion density gradient as a function of normalized poloidal flux $\psi/\psi_w$ for various island widths ($w/a=0$ to $0.4$). Increasing island width leads to localized flattening of the gradient near the resonant surface, suppressing the ITG drive. 
    (b) Corresponding linear growth rates of toroidal ITG modes for different island widths.}
    
   \label{density}
\end{figure*}

\begin{figure*}
    \centering
    \includegraphics[scale=0.5]{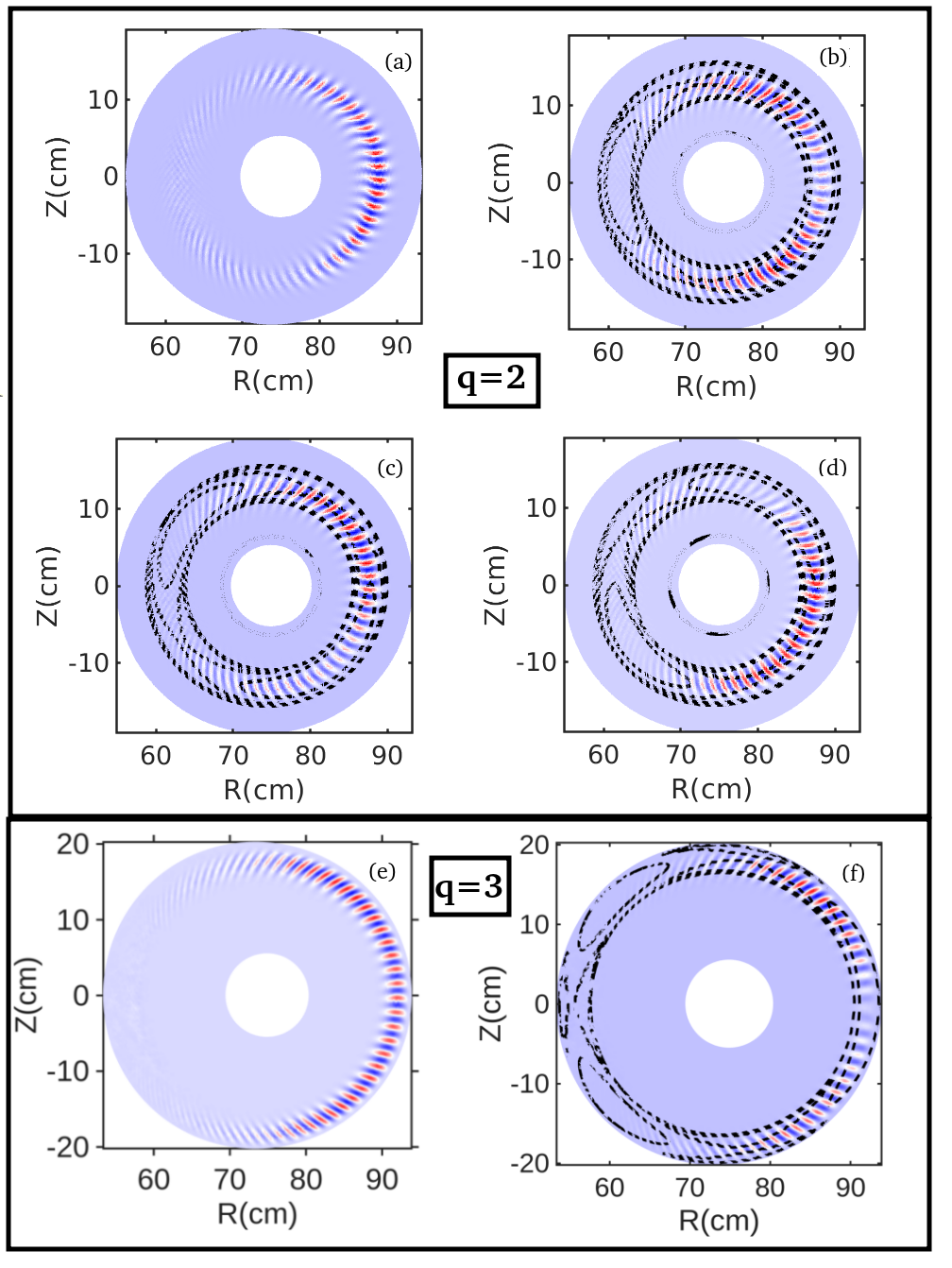}
    \caption{Electrostatic potential perturbation of ITG ($\delta\phi$) for the $(m,n) = (2,1)$ and $(3,1)$ magnetic-island cases at different poloidal cross-sections. 
    (a) Without island; (b-d) With an island of width $w = 0.29a$ (where $a$ is the minor radius) at different toroidal angles: 
    (b) $\zeta = 0$; (c) $\zeta = -\pi/4$; (d) $\zeta = \pi/4$; 
    (e) Without island for the $q = 3$ case; and (f) $\zeta = 0$ for the $q = 3$ island. 
    In the presence of the magnetic island, the ITG mode exhibits a pronounced asymmetry: the potential fluctuation amplitude is strongly localized near the island boundary and suppressed at the island center, reflecting the modified equilibrium gradients and the field-line topology induced by the island.}
    \label{mode}
\end{figure*}

\section{Results}\label{sec:results}

\noindent
 Fig.~\ref{density}(a) presents the radial ion density gradient profiles for various magnetic island widths, plotted against the normalized poloidal flux $\psi/\psi_w$. In the absence of an island ($w = 0$), the density gradient remains negative throughout the core, serving as the primary driver for the ion temperature gradient (ITG) instability. As island width increases from $w/a = 0.2$ to $0.4$, a region of nearly zero density gradient develops around the resonant surface. This flattening occurs because particles move much more easily along magnetic field lines than across them; their motion rapidly redistributes density along the closed field lines within the island, leading to a nearly uniform density profile inside the island region. In addition to this flattening, the location of the steepest remaining gradient shifts slightly inward. This inward shift occurs because the island-induced relaxation extends farther toward the outer side of the resonant surface, where magnetic field lines are longer, and particle motion along them is more effective at equalizing density. As a result, the outer gradient weakens more strongly, and the residual gradient region is displaced toward the high-field side. Together, these effects establish a new equilibrium configuration.

 \noindent Fig.~\ref{density}(b) shows that the ITG growth rate ($\gamma$) converges to the mean value of $0.432$ with the standard deviation of $0.019$, which shows that $\gamma$ is independent of the island width.  To understand how these profile modifications influence the mode structures, Fig.~\ref{mode}{(a)-(d)} provides poloidal snapshots of the electrostatic potential fluctuation $\delta\phi$ at $\zeta = 0$ and $\zeta = \pm \pi/4$ for $w/a = 0.29$ for a $q=2$ which clearly shows toroidal-angle dependence: the mode amplitude is suppressed at the island center and strongly localized near the island boundary. The localization becomes asymmetric with respect to the poloidal angle, with the strongest enhancement appearing on the outboard, unfavorable-curvature side for $\zeta=0$. At other toroidal phases, the regions of maximum potential fluctuation shift poloidally in accordance with the helical deformation imposed by the island geometry, rather than remaining fixed on the low-field side. In contrast, the reference case without an island, shown in Fig.~\ref{mode}(a), exhibits a more poloidally symmetric mode structure, confirming that the presence of a magnetic island fundamentally reshapes the spatial distribution of potential fluctuations.

 \noindent For the $q=3$ island shown in Fig.~\ref{mode}{(e)-(f)}, the overall potential fluctuation pattern retains the main qualitative features observed in the q=2 case but exhibits a noticeably broader spatial distribution. This broader structure arises from geometric effects: at higher-q surfaces, magnetic field lines are more tightly wound and have longer connection lengths $(L_{c}\sim qR)$\cite{Wesson2011}, allowing perturbations to extend over a larger region along the field.

\noindent Taken together, these results demonstrate that static magnetic islands modify the ITG instability through two complementary mechanisms. The first is equilibrium-profile flattening, which reduces local density and temperature gradients, thereby weakening the free-energy source driving the instability within the island region. The second is the modification of the magnetic field-line topology, which reshapes the ITG modes and redistributes potential fluctuation structures toward the island boundary. Both effects become more pronounced as the island width increases, since a wider island perturbs a larger fraction of the flux surface. However, the overall influence of the island diminishes at higher safety factors(q). At larger q, magnetic field lines wind around the torus more tightly, resulting in longer connection lengths. As a consequence, perturbations remain coupled over greater distances along the field, allowing potential fluctuations to spread more evenly and reducing their localization within the island region. In addition, the magnetic curvature drive, which provides the free energy for the ITG instability, becomes weaker at larger radii. Together, these geometric effects make higher-q regions less susceptible to island-induced distortion of the ITG mode structure, thereby reducing the impact on linear growth rates. Overall, these findings highlight the strong sensitivity of ITG instability to magnetic topology and demonstrate that island-induced equilibrium modifications act as a stabilizing influence by simultaneously weakening the linear drive and redistributing potential fluctuation activity within and around the island region, thereby improving local stability.

\section{conclusion}\label{sec:conclusion}

\noindent In this study, we investigate the influence of static magnetic islands on ion temperature gradient (ITG) instabilities relevant to the ADITYA-U tokamak using the Global Gyrokinetic Code in Cylindrical Coordinates (G2C3). The code’s neural–network–assisted particle projection method enables accurate treatment of field-aligned dynamics in the presence of island-induced magnetic topology, avoiding numerical difficulties near the island boundary. Static magnetic islands with mode numbers $(m,n)=(2,1)$ and $(3,1)$, experimentally observed in ADITYA-U \cite{Tanna2019, Tanna2024, Macwan2021DTM}, are introduced in the simulations by perturbing the equilibrium magnetic flux functions, corresponding to the $q=2$ and $q=3$ resonant surfaces. Particle-in-cell simulations in these modified magnetic configurations show clear density flattening within the island regions, confirming island formation and its impact on the equilibrium profiles via field-line reconnection. Linear electrostatic gyrokinetic simulations performed on these relaxed equilibria reveal that magnetic islands significantly modify ITG dynamics. In both island cases, the ITG mode exhibits reduced potential fluctuation activity within the island core and enhanced localization toward the island boundary, reflecting the altered equilibrium gradients and magnetic curvature distribution. As the island width increases, the growth rates of different toroidal ITG modes converge, indicating that sufficiently large islands reduce the sensitivity of the instability to mode number and radial localization. A comparison between the $q=2$ and $q=3$ cases further shows that higher-q islands give rise to potential fluctuation patterns that are more spatially extended along the magnetic field lines. This broader spatial structure is consistent with the longer connection lengths and weaker curvature effects at larger radii, which allow perturbations to spread over a wider region surrounding the island.

\noindent Overall, these results demonstrate that magnetic islands exert a stabilizing influence on ITG instabilities through a combination of equilibrium profile flattening and spatial redistribution of potential fluctuation structures around the island region. These findings advance the understanding of island–microinstability interaction and provide a physics-based foundation for future studies of island control and confinement optimization in fusion devices such as ADITYA-U

\section*{Acknowledgements}
\noindent This work is supported by the Board of Research in Nuclear Sciences (BRNS Sanctioned no. and 57/14/04/2022-BRNS), Science and Engineering Research Board EMEQ program (SERB sanctioned no. EEQ/2022/000144), and National Supercomputing Mission (NSM). We acknowledge National Supercomputing Mission (NSM) for providing computing resources of `PARAM PRAVEGA' at S.E.R.C. Building, IISc Main Campus, Bangalore, which is implemented by C-DAC and supported by the Ministry of Electronics and Information Technology (MeitY) and Department of Science and Technology (DST), Government of India, and  ANTYA cluster at Institute of Plasma Research, Gujarat.

\bibliography{mybib}

@PREAMBLE{
 "\providecommand{\noopsort}[1]{}" 
 # "\providecommand{\singleletter}[1]{#1}%" 
}

@article{Tanna2019,
  author       = {R. L. Tanna and R. Jha and S. C. Sharma and others},
  title        = {Overview of operation and experiments in the ADITYA-U tokamak},
  journal      = {Nuclear Fusion},
  volume       = {59},
  number       = {11},
  pages        = {112006},
  year         = {2019},
  doi          = {10.1088/1741-4326/ab3f8c}
}

@article{hender2007mhd,
  title={MHD stability, operational limits and disruptions},
  author={Hender, TC and Wesley, JC and Bialek, J and Bondeson, A and Boozer, AH and Buttery, RJ and Garofalo, A and Goodman, TP and Granetz, RS and Gribov, Y and others},
  journal={Nuclear fusion},
  volume={47},
  number={6},
  pages={S128},
  year={2007},
  publisher={IOP Publishing}
}

@article{patel2023role,
  title={Role of magnetohydrodynamic activity in sawtooth induced heat pulse propagation in ADITYA tokamak},
  author={Patel, S and Ghosh, J and Chowdhuri, MB and Mayya, KBK and Manchanda, R and Pandya, HKB and Tanna, RL and Kumar, V and Joisa, S and Purohit, S and others},
  journal={Nuclear Fusion},
  volume={63},
  number={3},
  pages={036001},
  year={2023},
  publisher={IOP Publishing}
}

@article{wilson2009influence,
  title={The influence of magnetic islands on drift mode stability in magnetized plasma},
  author={Wilson, HR and Connor, JW},
  journal={Plasma Physics and Controlled Fusion},
  volume={51},
  number={11},
  pages={115007},
  year={2009},
  publisher={IOP Publishing}
}

@article{hornsby2011interaction,
  title={Interaction of turbulence with magnetic islands: effect on bootstrap current},
  author={Hornsby, WA and Siccinio, M and Peeters, AG and Poli, E and Snodin, AP and Casson, Francis James and Camenen, Y and Szepesi, G},
  journal={Plasma Physics and Controlled Fusion},
  volume={53},
  number={5},
  pages={054008},
  year={2011},
  publisher={IOP Publishing}
}

@article{Macwan2021DTM,
  author    = {Macwan, Tanmay and Pradhan, Subrata and Ghosh, J. and Tanna, R. L. and others},
  title     = {Controlling the rotation of drift-tearing modes by biased electrode in ADITYA-U tokamak},
  journal   = {Physics of Plasmas},
  volume    = {28},
  number    = {11},
  pages     = {112111},
  year      = {2021},
  doi       = {10.1063/5.0066347}
}

@article{Singh2023,
  author = {Singh, Tajinder and Sharma, Deepti and Macwan, Tanmay and Sharma, Sarveshwar and Ghosh, Joydeep and Sen, Abhijit and Lin, Zhihong and Kuley, Animesh},
  title = {Gyrokinetic simulations of electrostatic microturbulence in ADITYA-U tokamak},
  journal = {Nuclear Fusion},
  volume = {63},
  number = {5},
  pages = {056008},
  year = {2023},
  month = {mar},
  publisher = {IOP Publishing},
  doi = {10.1088/1741-4326/acc344},
  url = {https://iopscience.iop.org/article/10.1088/1741-4326/acc344}
}

@article{Singh_2024,
doi = {10.1088/1741-4326/ad5a20},
url = {https://doi.org/10.1088/1741-4326/ad5a20},
year = {2024},
month = {jun},
publisher = {IOP Publishing},
volume = {64},
number = {8},
pages = {086038},
author = {Singh, Tajinder and Shah, Kajal and Sharma, Deepti and Ghosh, Joydeep and Jadeja, Kumarpalsinh A. and Tanna, Rakesh L. and Chowdhuri, M.B. and Lin, Zhihong and Sen, Abhijit and Sharma, Sarveshwar and Kuley, Animesh},
title = {Gyrokinetic simulations of electrostatic microturbulence in ADITYA-U tokamak with argon impurity},
journal = {Nuclear Fusion},
}

@article{Li2023,
  title={Global gyrokinetic simulations of the impact of magnetic island on ion temperature gradient driven turbulence},
  author={Li, Jingchun and Xu, JQ and Qu, YR and Lin, Z and Dong, JQ and Peng, XD and Li, JQ},
  journal={Nuclear Fusion},
  volume={63},
  number={9},
  pages={096005},
  year={2023},
  publisher={IOP Publishing}
}

@article{Tanna2024,
  author = {Tanna, R. L. and others},
  title = {Overview of physics results from the ADITYA-U tokamak and future experiments},
  journal = {Nuclear Fusion},
  volume = {64},
  number = {11},
  pages = {112011},
  year = {2024},
  publisher = {IOP Publishing},
  doi = {10.1088/1741-4326/ad4f22},
  url = {https://www.researchgate.net/publication/379702028_Overview_of_physics_results_from_the_ADITYA-U_tokamak_and_future_experiments}
}

@article{ F.L.,
  title={Theory and observations of magnetic islands},
  author={Waelbroeck, FL},
  journal={Nuclear Fusion},
  volume={49},
  number={10},
  pages={104025},
  year={2009},
  publisher={IOP Publishing}
}

@article{ Yamada,
  title={Magnetic reconnection},
  author={Yamada, Masaaki and Kulsrud, Russell and Ji, Hantao},
  journal={Reviews of modern physics},
  volume={82},
  number={1},
  pages={603--664},
  year={2010},
  publisher={APS}
}

@article{ Harold,
  title={Finite-resistivity instabilities of a sheet pinch},
  author={Furth, Harold P and Killeen, John and Rosenbluth, Marshall N},
  journal={The physics of Fluids},
  volume={6},
  number={4},
  pages={459--484},
  year={1963},
  publisher={AIP Publishing}
}

@article{ Fitzpatrick,
  title={The influence of the ion polarization current on magnetic island stability in a tokamak plasma},
  author={Fitzpatrick, R and Waelbroeck, FL and Militello, F},
  journal={Physics of plasmas},
  volume={13},
  number={12},
  year={2006},
  publisher={AIP Publishing}
}

@article{ choi,
  title={Effects of plasma turbulence on the nonlinear evolution of magnetic island in tokamak},
  author={Choi, Minjun J and Bard{\=o}czi, L{\=a}szlo and Kwon, Jae-Min and Hahm, TS and Park, Hyeon K and Kim, Jayhyun and Woo, Minho and Park, Byoung-Ho and Yun, Gunsu S and Yoon, Eisung and others},
  journal={Nature communications},
  volume={12},
  number={1},
  pages={375},
  year={2021},
  publisher={Nature Publishing Group UK London}
}

@book{Wesson2011,
  author = {Wesson, J.},
  title = {Tokamaks},
  edition = {4},
  publisher = {Oxford University Press},
  year = {2011},
  isbn = {9780199592234}
}

@article{LaHaye2006,
  author = {La Haye, R. J.},
  title = {Neoclassical tearing modes and their control},
  journal = {Physics of Plasmas},
  volume = {13},
  number = {5},
  pages = {055501},
  year = {2006},
  doi = {10.1063/1.2180747}
}

@article{Tanna2018,
  author = {Tanna, R. L. and Ghosh, J. and Chattopadhyay, P. K. and Raj, H. and Patel, S. and Dhyani, P. and Gupta, C. N. and Jadeja, K. A. and Patel, K. M. and Bhatt, S. B. and others},
  title = {Overview of recent operation and experiments  experiments in ADITYA and ADITYA-U tokamak},
  journal = {Nuclear Fusion},
  volume = {58},
  number = {10},
  pages = {106013},
  year = {2018},
  doi = {10.1088/1741-4326/aad3c7}
}

@article{Wei2025KSTAR,
  author    = {Wei, Xishuo and Nicolau, Javier H and Choi, Gyungjin and Lin, Zhihong and Yang, Seong-Moo and Kim, SangKyeun and Lee, WooChang and Zhao, Chen and Cote, Tyler and Park, JongKyu and Orlov, Dmitri},
  title     = {Gyrokinetic simulations of the effects of magnetic islands on microturbulence in {KSTAR}},
  journal   = {Nuclear Fusion},
  year      = {2025},
  volume    = {65},
  pages     = {026026},
  doi       = {10.1088/1741-4326/ada049},
  month     = {jan},
  note      = {Published 6 January 2025. Accepted for publication 17 December 2024. Received 30 September 2024.}
}

@article{De2019Kinetic,
  author    = {De, S. and Singh, T. and Kuley, A. and Bao, J. and Lin, Z. and Sun, G. Y. and Sharma, S. and Sen, A.},
  title     = {Kinetic particle simulations in a global toroidal geometry},
  journal   = {Physics of Plasmas},
  volume    = {26},
  number    = {8},
  pages     = {082507},
  year      = {2019},
  month     = {aug},
  doi       = {10.1063/1.5108684},
  url       = {https://doi.org/10.1063/1.5108684}
}

@article{Jiang2014,
  author    = {P. Jiang and Z. Lin and I. Holod and C. Xiao},
  title     = {Effects of magnetic islands on drift wave instability},
  journal   = {Physics of Plasmas},
  volume    = {21},
  number    = {12},
  pages     = {122513},
  year      = {2014},
  publisher = {AIP Publishing},
  doi       = {10.1063/1.4903910},
  url       = {https://doi.org/10.1063/1.4903910}
}

@article{alageshan2024neural,
  title={Neural network assisted electrostatic global gyrokinetic toroidal code using cylindrical coordinates},
  author={J. K. Alageshan and J. Das and T. Singh and S. Sharma and A. Kuley},
  journal={arXiv:2408.12851v1},  
  year={2024}
}

@article{Singh_2022,
doi = {10.1088/1741-4326/ac906d},
url = {https://doi.org/10.1088/1741-4326/ac906d},
year = {2022},
month = {oct},
publisher = {IOP Publishing},
volume = {62},
number = {12},
pages = {126006},
author = {Singh, Tajinder and Nicolau, Javier H. and Lin, Zhihong and Sharma, Sarveshwar and Sen, Abhijit and Kuley, Animesh},
title = {Global gyrokinetic simulations of electrostatic microturbulent transport using kinetic electrons in LHD stellarator},
journal = {Nuclear Fusion},
}

@article{Singh_2024a,
doi = {10.1088/1741-4326/ad0aca},
url = {https://dx.doi.org/10.1088/1741-4326/ad0aca},
year = {2023},
month = {nov},
publisher = {IOP Publishing},
volume = {64},
number = {1},
pages = {016007},
author = {Tajinder Singh and Javier H. Nicolau and Federico Nespoli and Gen Motojima and Zhihong Lin and Abhijit Sen and Sarveshwar Sharma and Animesh Kuley},
title = {Global gyrokinetic simulations of electrostatic microturbulent transport in LHD stellarator with boron impurity},
journal = {Nuclear Fusion}
}

@article{lee1987gyrokinetic,
  title={Gyrokinetic particle simulation model},
  author={Lee, WW},
  journal={Journal of Computational Physics},
  volume={72},
  number={1},
  pages={243--269},
  year={1987},
  publisher={Elsevier}
}
\bibliographystyle{unsrt}
\end{document}